\documentclass[aps,prc,floats,floatfix,preprint,superscriptaddress,nofootinbib]{revtex4}

\usepackage{graphicx}
\def\be{\begin{equation}}
\def\ee{\end{equation}}
\def\bea{\begin{eqnarray}}
\def\eea{\end{eqnarray}}
\def\rb{\rangle}
\def\lb{\langle}
\def\calA{$\cal A$}
\def\calB{$\cal B$}
\begin{document}

\title{Magnetic circular dichroism in real-time time-dependent density
functional theory}

\author{K.-M. Lee}
\affiliation{
Graduate School of Science and Technology, University of Tsukuba,
Tsukuba 305-8571, Japan}
\author{K. Yabana}
\affiliation{
Graduate School of Science and Technology, University of Tsukuba,
Tsukuba 305-8571, Japan}
\affiliation{
Center for Computational Sciences,
University of Tsukuba, Tsukuba 305-8571, Japan }
\author{G.F. Bertsch}
\affiliation{Institute for Nuclear Theory and Dept. of Physics,
University of Washington, Seattle, Washington}

\begin{abstract}

We apply the adiabatic time-dependent density functional theory to magnetic circular dichroism (MCD) spectra using
the real-space, real-time computational method.  The standard formulas for
the MCD response and its ${\cal
A}$ and ${\cal B}$ terms are derived from the observables in the
time-dependent wave function. 
We find real time method is well suited for calculating the overall 
spectrum, particularly at higher excitation energies where individual
excited states are numerous and overlapping.  The MCD sum rules
are derived and intepreted in the real-time formalism; we find that they
are very useful for normalization purposes and assessing the accuracy of the
theory.  The method is applied to MCD spectrum of C$_{60}$ using the
adiabatic energy functional from the local density approximation.  The 
theory correctly predicts the signs of the ${\cal
A}$ and ${\cal B}$ terms for the lowest allowed excitations.  However,
the magnitudes of the terms only show qualitative agreement with experiment.
\end{abstract}
\maketitle

\section{Introduction}

Magnetic circular dichroism (MCD) is an important spectroscopic observable
useful for characterizing the electronic structure of
molecules \cite{st76}  and condensed matter systems~\cite{we04}.
The theory of the MCD response
is usually given as a third-order perturbation in a basis that
diagonalizes the zero-field Hamiltonian.  This formulation,  called 
the sum-over-states method, requires a considerable computational
effort to construct the states and perform the summations.
There have been recent attempts to 
simplify the calculation by using eigenstates of the Hamiltonian in 
the presence of the magnetic field \cite{se08,so08}, but one still 
requires a sum over transition densities.
We propose here a completely different formalism based
on the solution of time-dependent equations of motion
and present a formalism for \calA~and \calB~terms of the MCD. 
We find that the formalism is a practical one when applied in the 
framework of time-dependent density functional theory (TDDFT).  
In fact, the TDDFT has already
been used successfully to calculate MCD in molecules \cite{se04,se07,kr07}.
A separate problem in the theory of MCD is the choice of a basis set
to construct the electron orbital wave functions.  The MCD puts higher
demands on the orbital representation to satisfy completeness and gauge
invariance.  In our treatment, we represent the
orbital wave functions on a spatial mesh rather than with atom-centered
basis set. The calculated matrix elements are automatically gauge-invariant
and one also avoids the inconsistencies that cause sum rules to be
violated.

We mention that the real-time TDDFT has been applied 
to many observables related to electron dynamics \cite{ya96, ya06, ca06}.
Specific applications include the molecular absorption spectrum in the
continuum \cite{na01},
hyperpolarizabilities \cite{iw04,ta07,an07}, 
the dielectric function \cite{be00},
and chiral dichroism \cite{ya99}.  
The real-time method has also been applied to phenonema associated with high
fields. In the presence of high fields, there is
hardly any alternative theory available, at least at the {\it ab initio}
DFT level.  Applications include nonlinear electron dynamics in metallic 
clusters \cite{ca00},
high harmonic generation \cite{no04}, Coulomb explosion \cite{ca04, ka09a}
dielectric breakdown \cite{ot08}, and coherent phonon generation \cite{sh10}. 

The organization of this article is as follows.  In Section II we define 
the time-dependent quantities that are computed and derive the
formulas for extracting the observables related to MCD.  
We also review the sum rules satisfied by the MCD response in that Section.  
In  Section III we provide some of the numerical details in carrying out
the MCD calculations.  Following that, in Section IV, we apply the
theory to the C$_{60}$ molecule. Due to its high symmetry, the C$_{60}$ 
molecule can exhibit both \calA~ and
\calB~terms of MCD.  The lowest electronic excitations of this
molecule are the $\pi-\pi^*$ character; for the measured transitions
we find the correct signs for the calculated $\cal A$ and $\cal B$ terms.  The
theory is in qualitative agreement also with magnitude of the \calB~term
of the lowest transition.  However, the present theory does not reproduce
well the other transitions and the magnitude of the \calA~term.

\section{Theory}

\subsection{Definitions for MCD response function}

We consider a molecule under a static magnetic field $B$ in
$\gamma$ direction. The electronic Hamiltonian is written as
\be
H = H_0 + \mu_B B L_{\gamma},
\ee
where $H_0$ is the Hamiltonian in the absence
of the magnetic field, $L_{\gamma}$ is the angular momentum 
operator, and $\mu_B = e/2mc$ is the Bohr magneton. 
We take a convention of $e>0$ and $\hbar=1$. 
We denote the ground and excited states under the static 
magnetic field $B$ as
\be
H \Phi_n = E_n \Phi_n.
\ee

We denote the dipole operator as $\vec \mu = -e \sum \vec r_i$
where $\vec r_i$ are electron coordinates.
We define the circularly polarized dipole operators with the 
normalization convention
$
\mu_\pm^{(\gamma)} = (\mu_\alpha\pm i \mu_\beta)/\sqrt{2},
$
where $(\alpha\beta\gamma)$ is a cyclic permutation of $(xyz)$.

In MCD, the basic object of study is the difference in dipole 
absorption strength functions for light of opposite circular 
polarization in a weak magnetic field. The MCD response may be 
defined by the strength function
\be
\label{def0}
R_{\rm MCD}(E) = \frac{1}{3 \mu_B B}
\sum_n \sum_\gamma
\delta( E - E_{n0}) \left\{
|\langle \Phi_0 | \mu_-^{(\gamma)} | \Phi_n \rangle|^2
- |\langle \Phi_0 | \mu_+^{(\gamma)} | \Phi_n \rangle|^2\right\},
\ee
where $E_{n0}$ is the excitation energy of state $n$,
$E_{n0}=E_n-E_0$.
The beam direction coincides with the magnetic field direction 
along an axis labeled by $\gamma$.  
It is convenient to express the strength function in a Cartesian 
basis using the antisymmetric tensor $\epsilon_{\alpha\beta\gamma}$, 
\be
\label{MCD}
R_{\rm MCD}(E) = 
-\frac{1}{3 \mu_B B}
\sum_{\alpha\beta\gamma} \epsilon_{\alpha\beta\gamma}
\sum_n \delta(E-E_{n0})
{\rm Im} \left\{ 
\langle \Phi_0 \vert \mu_{\alpha} \vert \Phi_n \rangle
\langle \Phi_n \vert \mu_{\beta} \vert \Phi_0 \rangle
\right\},
\ee
where the magnetic field is applied to $\gamma$ direction.

To assess the quality of the theory, it is also useful 
to calculate the ordinary dipole response. We define the 
dipole response $R_D(E)$
\be
R_D(E) = \frac{1}{3}\sum_n \sum_{\alpha} \delta(E-E_{n0})
| \langle \Phi_0 \vert \mu_{\alpha} \vert \Phi_n \rangle |^2,
\ee
This is related to the oscillator strength distribution by
\be
\label{dfdE}
\frac{df}{dE} = \frac{2m E_{n0}}{e^2} R_D(E).
\ee

Below the ionization threshold, electronic states are discrete.
In this energy region, the MCD strength function is often written as
\be
R_{\rm MCD}(E) = \sum_n \left\{
{\cal A}_n \left( -\frac{d}{dE} {\cal F}_n(E-E^{(B=0)}_{n0}) \right)
+ {\cal B}_n {\cal F}_n(E-E^{(B=0)}_{n0}) \right\},
\ee
where ${\cal F}_n(E)$ is the spectral shape of the  $n$-th state
normalized as $\int {\cal F}_n(E) dE = 1$.
The zero-field excitation energy is expressed as $E^{(B=0)}_{n0}$.
There appear both \calA~and \calB~terms for molecules with 
degeneracy in either ground or excited states, while only \calB~term 
appears for molecules without degeneracy in any states.
Integrating the MCD response function over an excitation energy
around $E^{(B=0)}_{n0}$, we have
\be
\label{An}
{\cal A}_n = \int^{E_{n0}+\epsilon}_{E_{n0}-\epsilon} d E
(E - E_{n0}^{(B=0)} ) R_{\rm MCD}(E),
\ee
and
\be
\label{Bn}
{\cal B}_n  = \int^{E_{n0}+\epsilon}_{E_{n0}-\epsilon} 
d E R_{\rm MCD}(E),
\ee
where $\epsilon$ is a small energy interval.

Employing the perturbation theory, these terms may be expressed in
terms of the energy and the wave functions in the absence of the 
magnetic field. For \calA, we have
\be
{\cal A}_n = -\frac{1}{3}\sum_{\alpha\beta\gamma} \epsilon_{\alpha\beta\gamma}
\sum_{st} \left\{ (L_{\gamma})_{nt,nt} - (L_{\gamma})_{0s,0s} \right\}
{\rm Im} \lb \Phi_{0s} | \mu^{(\gamma)}_{\alpha} | \Phi_{nt} \rb
         \lb \Phi_{nt} | \mu^{(\gamma)}_{\beta}  | \Phi_{0s} \rb,
\ee
where $s$ and $t$ distinguishes degenerate states of ground and excited 
states, respectively. The basis which diagonalize $L_{\gamma}$ is assumed. 
For \calB, we have
\bea
{\cal B}_n = \frac{2}{3} {\rm Im} 
\sum_{\alpha\beta\gamma} \epsilon_{\alpha\beta\gamma}
&&\sum_m \left\{ \frac{1}{E_{m0}}
\lb \Phi_m | L_{\gamma} | \Phi_0 \rb
\lb \Phi_0 | \mu_{\alpha} | \Phi_n \rb
\lb \Phi_n | \mu_{\beta}  | \Phi_m \rb \right. \nonumber\\
&&+
\left. \frac{1}{E_{mn}}
\lb \Phi_n | L_{\gamma} | \Phi_m \rb
\lb \Phi_0 | \mu_{\alpha} | \Phi_n \rb
\lb \Phi_m | \mu_{\beta}  | \Phi_0 \rb \right\}.
\eea

Similarly, we define the ordinary dipole strength as
\be
{\cal D}_n  = \int^{E_{n0}+\epsilon}_{E_{n0}-\epsilon} d E R_D(E)
= \frac{1}{3}\sum_{\alpha} | \lb \Phi_0 | \mu_{\alpha}  | \Phi_n \rb |^2.
\ee
It is related to the oscillator strength $f_n$ as
$f_n = 2 m E_{n0} {\cal D}_n/e^2$.  

Finally, with our definition of the MCD response, the \calB~coeffient
is related to the ratio of the MCD extinction coefficient $\Delta \epsilon$
to the ordinary extinction coefficient $\epsilon$ by the formula
\be
{\Delta \epsilon \over \epsilon} = {2 m \mu_B B E_{n0} {\cal B}_n\over e^2 f_n}%
\ee

\subsection{Real time formulation}
The response associated with any pair of operators,
\def\oa{{\cal O}_1}
\def\ob{{\cal O}_2}
$\oa$ and $\ob$, may be calculated in the real time domain starting from
the ground state wave function $\Phi_0$.  One first applies an
impulsive perturbation $\lambda \oa$  to obtain an initial
wave function $\Psi(t=0_+)$.  This is then evolved in time with
the exponentiated Hamiltonian operator,
\be
\Psi(t) = e^{-i H t}e^{ i\lambda \oa} \Phi_0.
\ee 
The real-time response $S_{21}(t)$ is given by the expectation
value of the operator $\ob$ in that state,
\be
S_{21}(t) = 
\langle\Psi(t)|\ob|\Psi(t)\rangle.
\ee
The linear response is evaluated by treating $\lambda$ as a small parameter
and taking the derivative $d S_{21}(t) / d \lambda$
at $\lambda = 0$.  
Depending on the operators $\oa$ and $\ob$, the strength function $R_{21}$ 
is obtained as the imaginary or real part of the
Fourier transform of linear response 
$d S_{21}(t) / d \lambda \vert_{\lambda=0} $ 
on the time interval $[0,+\infty]$.

This general formulation of the linear response applies to the MCD
strength function Eq. (\ref{MCD}) taking the operators to be
$ \oa=  \mu_\beta$  and $\ob = \mu_\alpha$.  The wave function is 
set up at $t=0$ as $\Psi_{k\beta}(t=0_+) = e^{ik \mu_\beta} \Phi_0$
and the real-time response $S_{\alpha\beta}^{(\gamma)}(t)$ is given by
\def\sabc{S_{\alpha\beta}^{(\gamma)}(t)}
\be
\sabc =
\lb \Psi_{k\beta}(t) | \mu_{\alpha} | \Psi_{k\beta}(t) \rb,
\ee
where $(\gamma)$ in $S_{\alpha\beta}^{(\gamma)}$ is included to 
remember that a static magnetic field is applied to $\gamma$ 
direction throughout the time evolution.
Expanding the perturbing operator $e^{i k \mu_\beta}$ in powers of 
$k$, we have
\bea
\sabc &=&
ik \left\{
\lb \Phi_0 | \mu_{\alpha} e^{-i(H-E_0)t} \mu_{\beta} | \Phi_0 \rb
-\lb \Phi_0 | \mu_{\beta} e^{i(H-E_0)t} \mu_{\alpha} | \Phi_0 \rb
\right\}. \nonumber\\
&=& ik \sum_n 
\lb \Phi_0 | \mu_{\alpha} | \Phi_n \rb 
\lb \Phi_n | \mu_{\beta} | \Phi_0 \rb
e^{-iE_{n0} t}
-ik \sum_n \lb \Phi_0 | \mu_{\beta} | \Phi_n \rb 
\lb \Phi_n | \mu_{\alpha} | \Phi_0 \rb
e^{iE_{n0} t}.
\eea
In the last formula, we have expressed the time-evolution operator in
terms of the energy eigenstates of the system.  We next separate
out the time-even and time-odd parts of the response, writing it as 
\bea
\label{Sabc}
\sabc
&=& -2k \sum_n {\rm Im}\,\left\{\lb \Phi_0 | \mu_{\alpha} | \Phi_n \rb 
\lb \Phi_n | \mu_{\beta} | \Phi_0 \rb \right\} \cos E_{n0} t \nonumber\\
&& +
2k \sum_n {\rm Re}\,\left\{\lb \Phi_0 | \mu_{\alpha} | \Phi_n \rb 
\lb \Phi_n | \mu_{\beta} | \Phi_0 \rb \right\}
\sin E_{n0} t.
\eea
\def\Sdiff{S_{\alpha\beta}^{(\gamma)}-S_{\beta\alpha}^{(\gamma)}}
The combination $\Sdiff= -4k \sum_n {\rm Im}\,\left\{\lb \Phi_0 | \mu_{\alpha} | \Phi_n \rb 
\lb \Phi_n | \mu_{\beta} | \Phi_0 \rb \right\} \cos E_{n0} t \nonumber\\
$ isolates the
first term with the cosine dependence on time.  We obtain an
expression proportional to the MCD response by taking its cosine Fourier
transform, 
\be
-\frac{1}{2\pi k} \int_0^{\infty} dt \cos E t
\left\{
\Sdiff
\right\}
= \sum_n 
{\rm Im}\,\left\{\lb \Phi_0 | \mu_{\alpha} | \Phi_n \rb 
\lb \Phi_n | \mu_{\beta} | \Phi_0 \rb 
\right\} \delta (E - E_{n0}).
\ee
The MCD strength function with the proper prefactor is given by the
integral over the real-time response as
\be
\label{Reps}
R_{\rm MCD}(E) = 
\frac{1}{6\pi \mu_B B k}
\sum_{\alpha\beta\gamma}
\epsilon_{\alpha\beta\gamma}
\int_0^{\infty} dt \cos Et
\left\{ 
\Sdiff
\right\}.
\ee
For the molecules we treat here, we can choose coordinate systems such that
the off-diagonal
response is antisymmetric, $\sabc= -S_{\beta\alpha}^{(\gamma)}(t)$.  
Then the second term in Eq. (\ref{Sabc}) is identically zero and 
Eq. (\ref{Reps}) reduces to
\be
\label{Rmcd}
R_{\rm MCD}(E) = \frac{2}{3\pi \mu_B B k}
\int_0^{\infty} dt \cos Et
\left\{ S^{(z)}_{xy}(t) + S^{(x)}_{yz}(t) + S^{(y)}_{zx}(t) \right\}.
\ee
This is our main result that will be applied to calculate the MCD.

For most if not all MCD spectra, the sign of $R_{\rm MCD}(E)$ 
on the infrared side of the lowest optical excitation is 
negative.  We shall call this the ``normal" sign.

The ordinary dipole response $R_D(E)$ may also be computed 
in the formalism as Fourier sine transform,
\be
\label{dipole}
R_D(E)=\frac{1}{3\pi k} \sum_{\alpha} 
\int_0^\infty d t\, \sin E t \,
S_{\alpha\alpha}^{(\gamma)}(t).
\ee 
Note that this can easily be evaluated at the same time as 
$R_{\rm MCD}(E)$ because the same time-dependent 
wave function is used in both.

\subsection{Sum rules}
The real-time formalism is very convenient for evaluating and 
verifying energy-weighted sum rules.  In particular, the MCD
response satisfies a quadratic sum rule that depends only on
the magnetic field strength and the number of electrons \cite{ca77}.
The connection to the time-dependent response may be easily 
derived by expanding $S_{\alpha\beta}^{(\gamma)}$
as a power series in time $t$,
\be
\label{s-t2}
\sabc
\simeq
s_0 + s_2 t^2 + \cdots.
\ee
Only even powers of $t$ are present in the series expansion, due to 
the suppression of the second term in Eq. (\ref{Sabc}).
The coeffients $s_0$ and $s_2$ can be readily expressed as commutators
of the $\mu$ operators and the Hamiltonian and evaluated analytically.
One finds
\be
s_0 = -2k {\rm Im} \sum_n \left\{
\lb \Phi_0 | \mu_{\alpha} | \Phi_n \rb 
\lb \Phi_n | \mu_{\beta} | \Phi_0 \rb \right\}
= ik \lb \Phi_0 | [\mu_{\alpha}, \mu_{\beta}] | \Phi_0 \rb = 0,
\ee
and 
\be
\label{S2t}
s_2 = k {\rm Im} \sum_n \left\{
\lb \Phi_0 | \mu_{\alpha} | \Phi_n \rb 
\lb \Phi_n | \mu_{\beta} | \Phi_0 \rb \right\}
E_{n0}^2
=
-\frac{ik}{2} \lb \Phi_0 |
[\mu_{\alpha},[H,[H,\mu_{\beta}]]] | \Phi_0 \rb
= \pm \frac{e^3 Bk}{2m^2c}N_e
\ee
where $N_e$ is the number of electrons and the sign reflects
the order of $(\alpha\beta\gamma)$.
These formulas can be expressed as integrals over the MCD strength function
\be
\label{S0}
I_0=\int_0^{\infty} dE R_{\rm MCD}(E) = 0
\ee
and
\be
\label{S2}
I_2=\int_0^{\infty} dE E^2 R_{\rm MCD}(E)
=
\frac{2e^2}{m}N_e.
\ee

The $I_2$ sum rule has a simple physical interpretation.  
Consider the real-time response associated with $S^{(z)}_{xy}$.
The impulsive exciting potential $e k y\, \delta(t)$  gives 
the electrons an average momentum equal to $-ek\hat y$, 
the integral of the force over time.  
The corresponding velocity, $ \vec v =-e k\hat y / m$, is subject to
a magnetic force $-e \vec v \times \vec B /c$ which is in the $x$ direction
for our geometry.  The corresponding acceleration is 
$a = e^2 B k\hat x/m^2c $.  Thus the $x$-component of the dipole 
moment increases quadratically with time according to the 
acceleration formula
\be
\label{st-t2}
\langle -e x(t)\rangle = {1\over 2} a t^2 = {N_e e^3 B k\over 2 m^2 c }t^2, 
\ee  
in agreement with Eq. (\ref{s-t2}) and (\ref{S2t}).

In the results of the calculations given below, we will also show the performance of the
theory with respect to the $f$-sum rule.  In terms of the
response $R_D(E)$, we define the quantity
\be
\label{ffE}
f_E = {2m}\int^E_0 d E' E' R_D(E').
\ee
The sum rule is given by 
\be
f_{+\infty} = N_e
\ee
where $N_e$ is the number of electrons.  The associated short-time
behaviour of the real-time response is linear in $t$ and given by
\be
\label{dipole-st}
\langle \Psi_{k\alpha}(t) | \mu_\alpha | \Psi_{k\alpha} (t) \rangle
\approx {e^2 k \over m} N_e t.
\ee

\subsection{Time-dependent density functional theory}

The basic variables in Kohn-Sham density functional theory are the
orbitals $\phi_i(\vec r)$, which are varied to minimize an energy expression
$E_{KS}$ that contains the quantum kinetic operator and terms depending on the
density $n(\vec r)= \sum_i |\phi_i(\vec r)|^2$.  The formal variation of the
$E_{KS}$ energy expression with respect to an orbital gives the Kohn-Sham
operator $H_{KS}$.  In the time-dependent extension of DFT with the
adiabatic approximation, the operator $H_{KS}$ also used in the
equation of motion for the orbitals. To linear order in the magnetic
field strength $B$, the time-dependent orbitals satisfy the equations
\be
\left[H_{KS} + \mu_B \vec L \cdot  \vec B\right]\psi_i(t) = i {\partial \over \partial t} \psi_i(t).
\ee
These equations are solved for  
$\psi_i(t)$ with initial condition 
$\psi_i(t=0_+) = e^{ik\mu_\beta} \phi_i(\vec r)$.
where $\phi_i$ are the ground-state Kohn-Sham orbitals.
The time-dependent response is calculated as
\be
\sabc = \sum_i \langle \psi_i(t) | \mu_\alpha | \psi_i(t)\rangle.
\ee
For the calculations described below, we treat only the valence 
electrons dynamically and ignore any spin dependence.  The effects of 
core electrons are treated by using Troullier-Martins pseudopotentials
for the electron-ion interaction \cite{tr91}.  We use the usual
LDA functional \cite{pe81} for the exchange-correlation interaction as in
previous work \cite{ka09b}.

\section{Computational Aspects}
The implementation of our real-time TDDFT is described in detail in 
Ref. \cite{ya06}.  An important difference from other TDDFT codes is
the orbital representation in 3-dimensional Cartesian
mesh.  This has the computational benefit that the Kohn-Sham operator
is given by a sparse matrix.  The representation is also convenient 
for treating extended wave functions such as Rydberg states or 
continuum states. It has
the disadvantage, however, that it does not permit all-electron calculations 
with practical mesh sizes.  For checking our code, we found it helpful
to calculate the observables with the Troullier-Martins pseudopotential
replaced by an anisotropic harmonic oscillator potential.  All the 
observables in this model have analytic expressions that can be compared
with the calculated numerical quantitites.  See the Appendix for details.

In our implementation of the mesh representation, the momentum operator 
$p$ is computed by 
the 8-point difference function, which is consistent with
our treatment of the kinetic operator $p^2/2m$ as a sum of 9-point difference
functions along the three Cartesian axes.  
The main numerical parameters in the calculation are the
mesh spacing $\Delta x$, the size of the spatial domain on
which the orbital wave functions are defined, 
the time step $\Delta t$, and the
total integration time $T$.  We find that adequate precision
for our purposes is obtained with parameter values
$\Delta x = 0.5$  and $\Delta t = 0.03$ in atomic units.  
The spatial dimensions needed for the orbital wave functions
depend on the desired accuracy in the continuum region.  The
continuum strength functions are smooth only if the spatial
domain is large and absorbing boundary conditions are applied
at the edges.  Typically, we take a cubical box of $160^3$ mesh points
for the calculations.  For small molecules, a much smaller domain is
adequate if the details of the response in the continuum are not
needed.

Although the TDDFT is fundamentally nonperturbative, the quantities
we calculate are in fact the perturbative limits with respect to
the strengths of the applied magnetic and electric fields.  We thus
choose strengths that are  small enough for the linear 
response formula to apply, but large enough to avoid numerical
roundoff errors.  For the perturbing electric field, we take 
$k=0.001$.  For the magnetic field, the calculations
reported below were carried out with a magnetic field 
given by $ \mu_B B= 0.0005$ au.  The intensity of this magnetic field is
0.137 au.  For comparison, a field strength of one Telsa has the value
$5.81 \times 10^{-4}$ in atomic units.

The integration time $T$ required to calculate the response
depends on the desired energy resolution.  We multiply the
integrands in the Fourier transforms Eq. (\ref{Reps}) and Eq. (\ref{dipole}) by
the filter function $1-3(t/T)^2+2(t/T)^3$ to smooth out spurious
oscillations from the upper time cutoff.  The resulting peaks 
associated with sharp states have a width $\Gamma$ 
(full width at half maximum) given 
approximately by $\Gamma\approx 6/T$.  Most of our results were
calculated by integrating $N_t=60000$ time steps, giving 
$\Gamma \approx 6/(N_t \Delta t)\sim 0.0033$ au $= 0.1$ eV. 

\section{Application to C$_{60}$}

The C$_{60}$ molecule offers a good test of the methodology to
demonstrate the feasibilty of using the 
real-time method as applied to fairly large molecules. 
Due to the high symmetry of C$_{60}$,
all optically allowed
transitions are three-fold degenerate and there will be both \calA~
and \calB~terms in the MCD spectrum.
There are 5-6 excitations in the calculated spectrum
up to 6 eV, all of which are $\pi-\pi^*$
character.  It has been found that the experimental oscillator 
strength \cite{ka08,ya09} accords well with the theory \cite{ka09b}
based on the {\it ab initio} adiabatic local density approximation.

\begin{figure}
\includegraphics [height = 6cm,viewport = 00 00 290 250,clip]{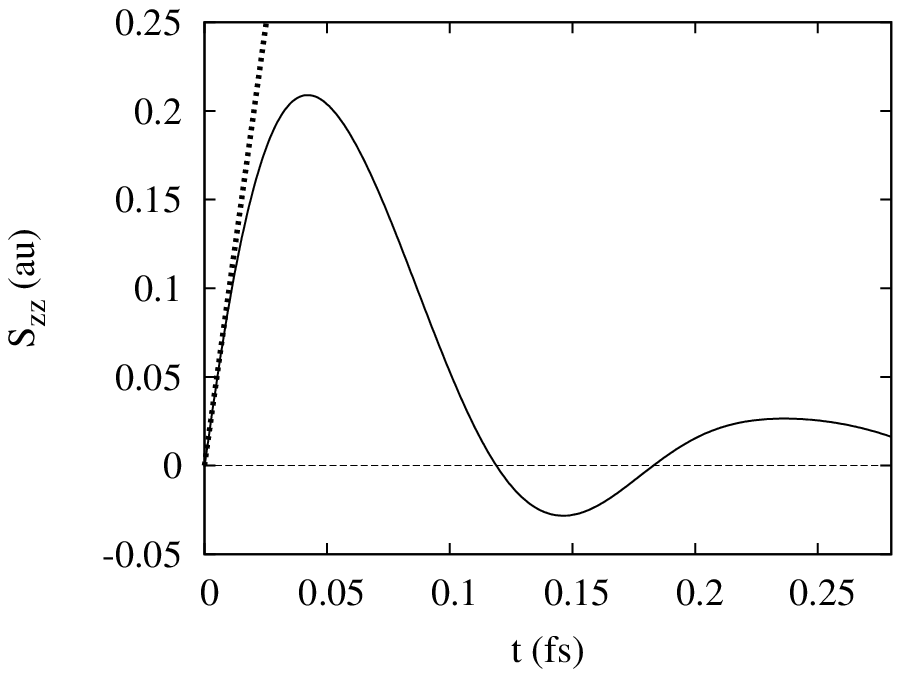}%
\includegraphics [height= 6cm,viewport = 00 00 350 250,clip]{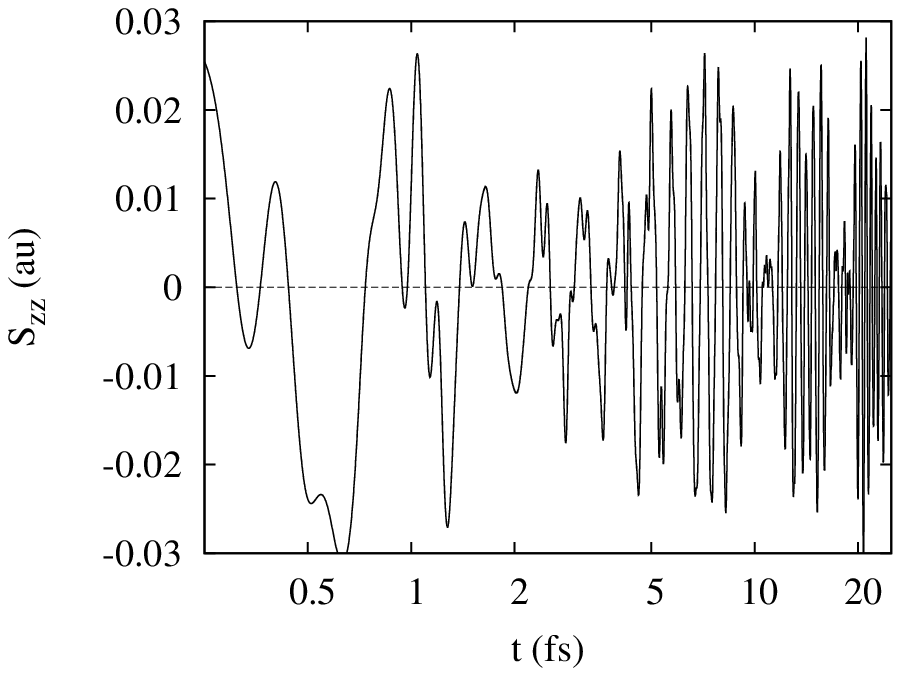}%
\caption{\label{rt_zz} Real-time dipole response 
$S_{zz}(t)=\langle \Psi_{kz}(t) | \mu_z | \Psi_{kz} (t) \rangle$
for C$_{60}$.  The left-hand panel shows the time interval $t=0-0.275$ fs with
a linear time scale. The sloping line shows the expected short-time 
behavior according to Eq. (\ref{dipole-st}).  
The right-hand panel shows the time
interval $t=0.25-25.0$ fs on a logarithmic time scale. 
}
\end{figure}

\begin{figure}
\includegraphics [width = 8 cm]{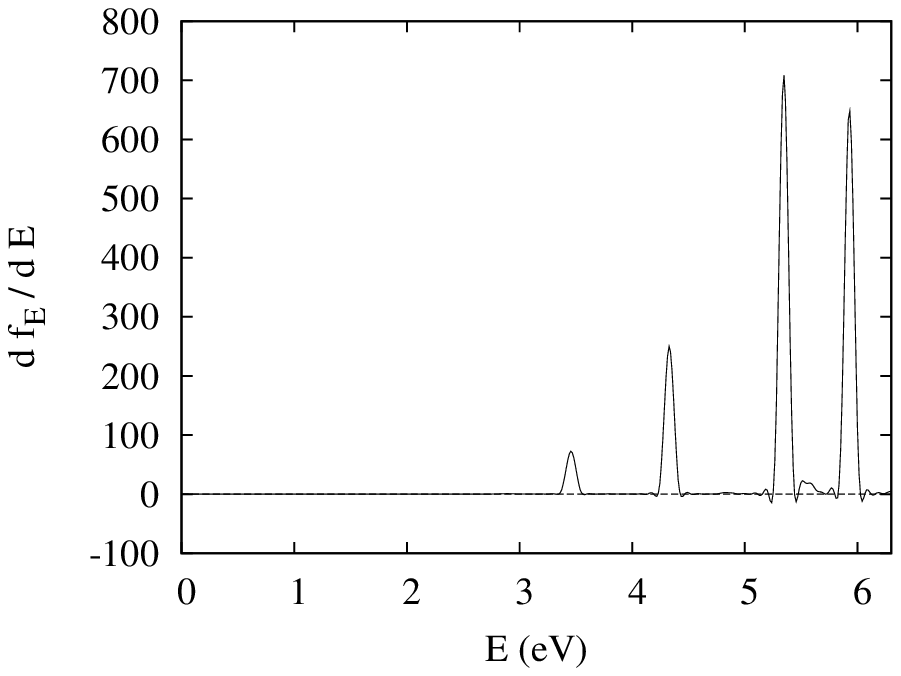}%
\includegraphics [width = 8cm]{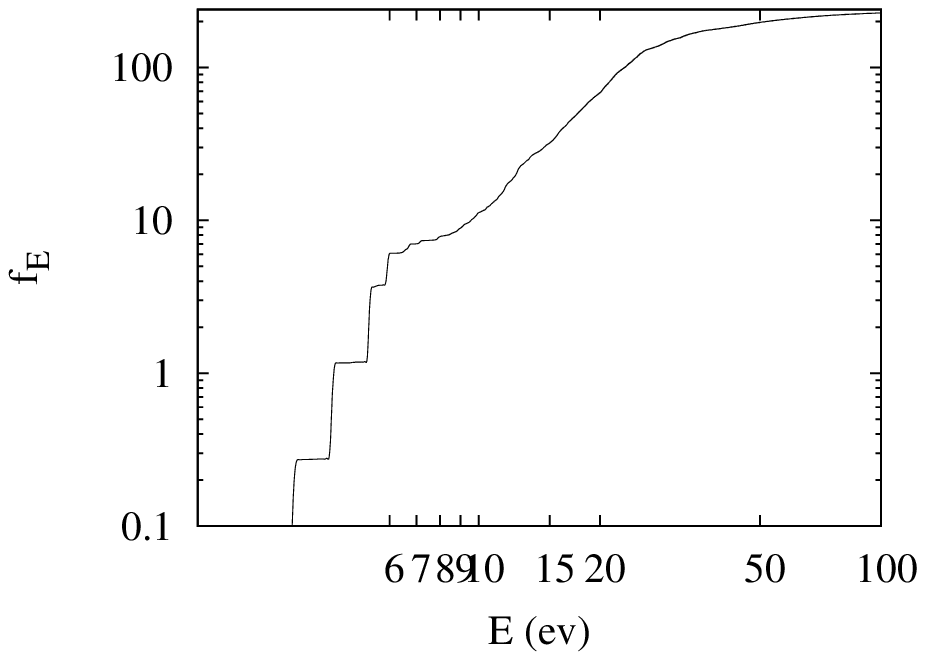}
\caption{\label{fig-fE} Dipole response for C$_{60}$.  The differential
oscillator strength $df/dE$ (Eq. (\ref{dfdE} )) up to 6 eV is shown in the
left-hand panel.  The right-hand panel shows the integrated oscillator
strength function $f_E$, Eq. (\ref{ffE}).}
\end{figure}

Our calculation here is very similar to that carried out in 
Ref. \cite{ka09b} for the oscillator strength function.  
The integration
time in the present calculation is somewhat longer,  60,000 time steps 
with $\Delta t = 0.03$ au.
Before examining the MCD response, we recall the results for the 
ordinary dipole response, as calculated in the real-time method.   
Fig.~\ref{rt_zz} shows the $S_{zz}(t)$ real-time
response over the interval $[0,T] = [0, 25]$ fs with the left-hand
panel showing an expanded view of the first 0.275 fs time interval.  
The short-time behavior expected from Eq. (\ref{dipole-st}) is shown by the
straight dotted line in the left-hand panel.   One may see that the
initial response does indeed follow Eq. (\ref{dipole-st}) very well.
After the initial rise in the first
0.1 fs the dipole moment oscillates 
with a period of order of one fs correspond to the strong transitions
in the energy interval 7-15 eV.  Note that the oscillation is essentially
undamped.  This is a consequence of the sharpness of the bound excitations
that would produce a $\delta$-function response if the Fourier transform 
could be done exactly.  The numerical Fourier transform was carried out to 
final time $t=1800$ au $= 43.5$ fs with the results for the low-frequency 
part of $R_D(E)$  shown in the left-hand
panel of Fig. \ref{fig-fE}.  There are four
transitions in the spectral region 0-6 eV, at excitation energies of 
3.5,4.3,5.3 and 5.9 eV.  The numerical FWHM widths are about 0.1 eV,
as expected from the integration time.
The important information besides the transition energy is
total strength in the individual peaks. This can be extracted from 
the graph of the integrated
strength $f_E$ defined in Eq. (\ref{ffE}).
The jumps at low energies give the $f$ strengths of the discrete
transitions. The total integrated
strength is $f=233$, rather close to the sum rule number $f=240$ for the
$N_e=240$ valence electrons treated dynamically in the TDDFT.
We note that the sum rule is not expected to be satisfied
exactly for our energy functional, because of nonlocality in the
Troullier-Martins pseudopotential.  

We now take up the MCD response.  The left-hand panel of 
Fig. 3 shows the calculated MCD real-time response 
$S^{(z)}_{xy}(t)$ over the time interval
$0 < t < 0.3$ fs.  The dashed curve in the left-hand 
panel shows the predicted short-time dependence according to 
Eq. (\ref{st-t2}). The computed time dependence starts out 
quadratic as expected, but the coefficient of $t^2$ 
is lower by 40\% than expected from Eq. (\ref{st-t2}).
To confirm
that the nonlocality of the pseudopotential is responsible for
the disagreement, we have recomputed the response for short times
with nonlocality of 
the pseudopotentials turned off, shown as the long-dashed line in 
the Figure.  This agrees closely with
the expected short-time behavior.
We do not have any explanation why the sum rule violation  is 
much stronger for the MCD strength than for the ordinary dipole strength.

\begin{figure}
\includegraphics [height = 6cm,viewport = 00 00 280 250,clip]{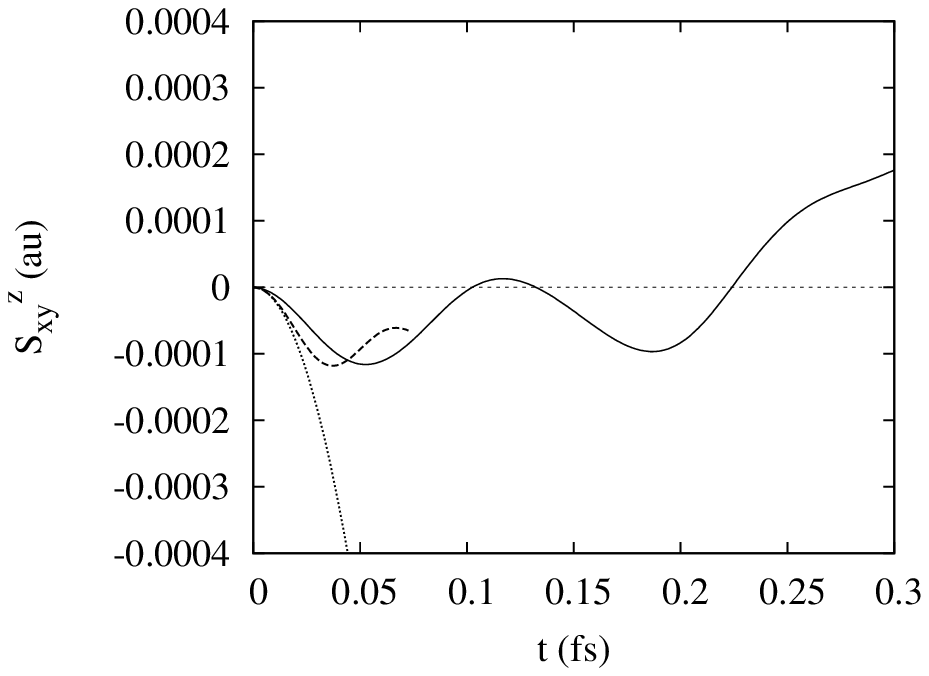}%
\includegraphics [height= 6cm,viewport = 00 00 350 250,clip]{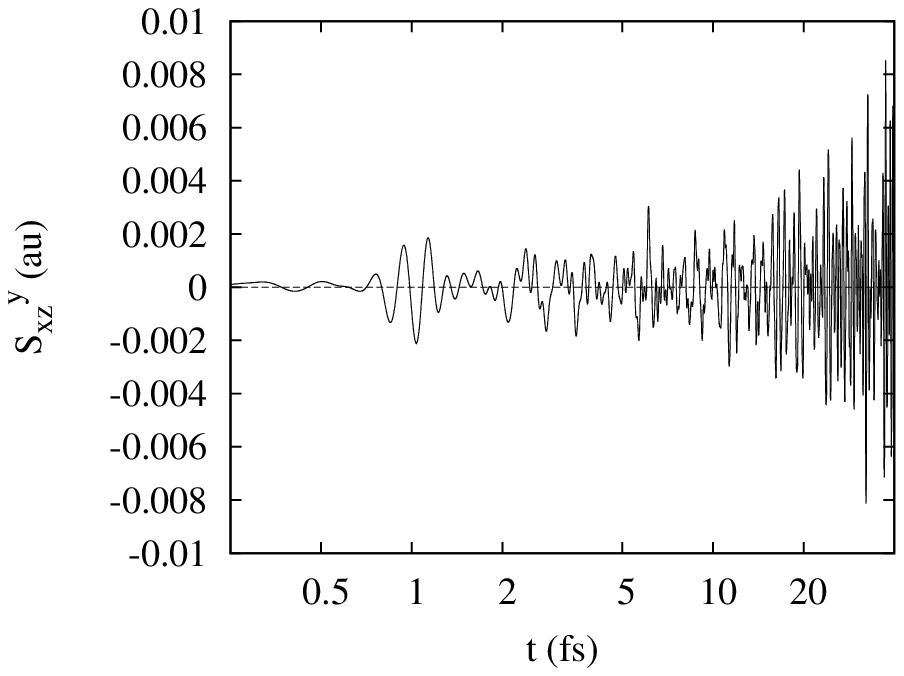}%
\caption{\label{c60-mcd} MCD real-time response $S^{(z)}_{xy} $ in
C$_{60}$. The solid line in left-hand panel shows the evolution for short
times, $0 < t < 0.3$ fs.  The
dotted curve is the expected dependence from Eq. (\ref{st-t2}).  
The long-dashed
curve shows the response in the time range $0 < t< 0.07$ fs with the
nonlocality in the pseudopotential turned off.
The right-hand panel shows the response in the longer 
time interval $0.25 < t < 40$ fs on a logarithmic time scale and a magnified
ordinate scale.
}
\end{figure}

\begin{figure}
\includegraphics[width = 9cm]{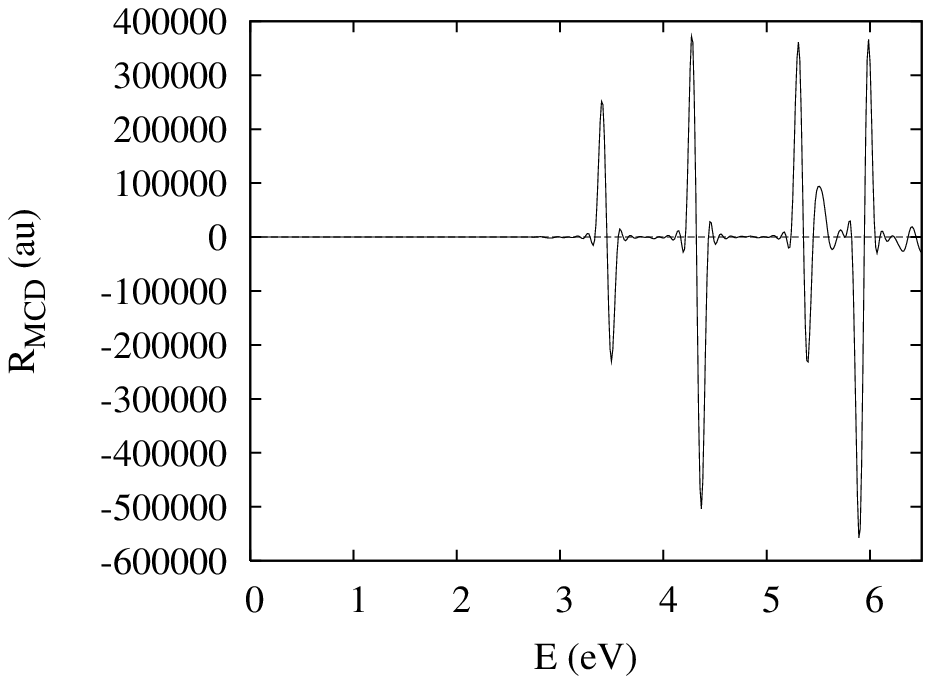}%
\includegraphics[width = 9cm]{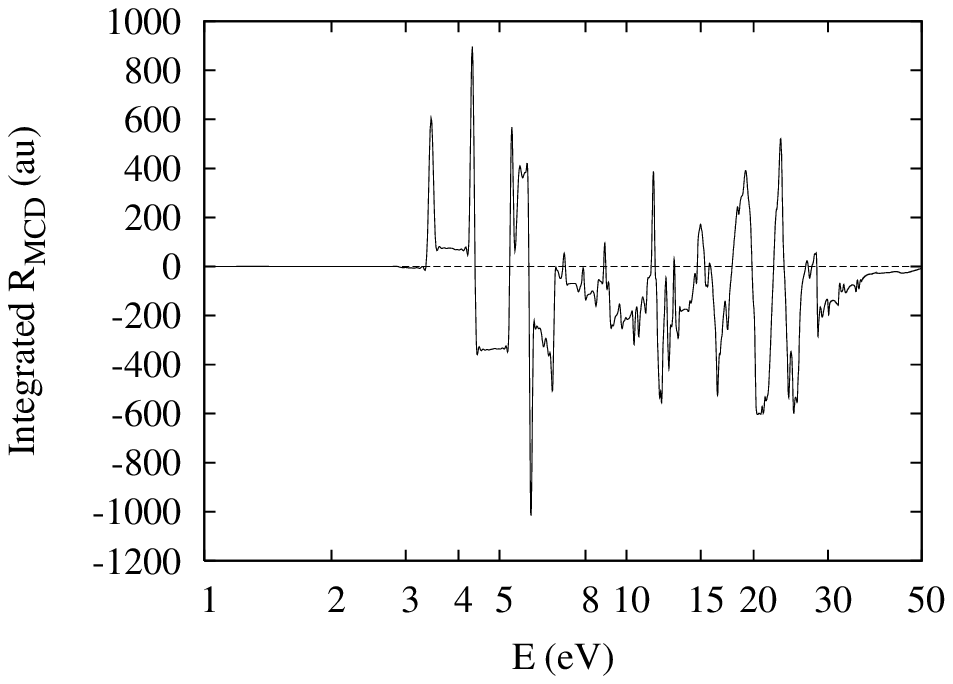}
\caption{\label{c60-mcd_e} MCD response $R_{\rm MCD}(E)$ in C$_{60}$.   
Left-hand panel shows the strength function Eq. (\ref{MCD}).  
The corresponding integrated strength function
 is shown in the right-hand panel.
}
\end{figure}

The MCD response going to long times is shown on the right-hand panel of 
Fig. \ref{c60-mcd}.  It is 
interesting to note that the amplitude of oscillation increases
with time.  This behavior is in contrast to the ordinary dipole response, 
which has a maximum excursion in the first oscillation. The reason for the
increase in amplitude is the presence of the \calA~terms which give a 
real-time response that cancels at short times and only becomes visible at
later times.
Taking the Fourier cosine transform of the real-time response using 
Eq.~(\ref{Reps}), we find the MCD spectrum shown in Fig. \ref{c60-mcd_e}, 
left-hand panel.  
The \calA-type character of the $\pi{-}\pi^*$ transitions is clearly seen in 
the shape of curves, each with a strong alternation of sign over the
width of the peaks in the dipole response function.  (Again, 
there is
no physical significance to the calculated widths since they depend
on the integration limit in the Fourier transform.)
It is interesting to see that the sign of \calA~coefficients can vary from
state to state. The excitation at 5.9 eV has the normal sign, namely 
negative on the low-frequency side, but the three lower excitations have
the opposite sign.  The four transitions in the figure also have a
significant \calB-type MCD response, visible by unequal positive- and 
negative-going peaks on the two sides of the transition.    
The \calB-type response may be seen more clearly in the graph of the integrated
MCD response, $\int^E d E' R_{\rm MCD}(E')$, shown on the right-hand 
panel of Fig.~\ref{c60-mcd_e}.  
The \calB$_n$~coefficients can be read off from the step increases
going across each transition, cf. Eq. (9).  The values are reported in 
Table I, divided by the theoretical dipoles strengths ${\cal D}_n$ (Eq. (12)).
This is to facilitate comparison to the experimental values \cite{ga91},
which are given in this form.  We see that the signs of ${\cal B}_n$ for
the lowest two states agree.  This is far from trivial.  Also, the
calculated magnitude of the lower one is within a factor of 2
of experiment.  This is poorer agreement than is typical for the
calculation of oscillator strength $f_n$ in TDDFT, but perhaps this
should not be unexpected due to the difficulties uncovered by the
unexpected short-time behavior.  Also, we know that there is considerable 
screening of the valence electron transition moments, amplifying
the relative errors of the screened observables.  The \calB$_n$ of the 
second state has a much larger discrepency.  Until that is understood,
one cannot use the TDDFT as a predictive tool for large molecules.  

For a overall view of the MCD response, the right-hand panel of 
Fig.~\ref{c60-mcd_e} shows the integrated
MCD response up to 50 eV.  The integrated response is
predominantly negative, as to be expected with the negative-going
initial evolution.  One sees that the total goes to zero at the upper energy,
showing that the $I_0$ sum rule (Eq. (26)) is
nearly satisfied.   Finally, the $E^2$ sum rule, Eq. (\ref{S2}),
has a value $I_2=258$, almost a factor of two smaller than the nominal
value of $2N_e=480$.  We have already seen this effect of the nonlocality
in the short-time response.

We next turn the \calA-type response, arising from
the energy splitting between members of 
$^1T_{1u}$ multiplets as in the Zeeman splitting. 
A convenient way to express the splitting is as the effective
$g$-factor for the transition \cite{st76}[Eq. (52)],
\be
\label{g}
g= {\Delta E\over \mu_B B}.
\ee
This is related to \calA$_n$ by
\be
g = \frac{{\cal A}_n}{2{\cal D}_n}
\ee
We extract the \calA$_n$~coefficients by Eq.~(\ref{An})
from $R_{\rm MCD}(E)$.
We may also extract the energy shift $\Delta E$ from the zero-field
value using the formula
\be
\Delta E = - \mu_B B{
\int^{E_{no}}_{E_{n0}-\epsilon} dE\, R_{\rm MCD}(E)-
\int^{E_{no+\epsilon}}_{E_{n0}} dE\, R_{\rm MCD}(E)\over 4 R_D(E_{n0})}
\ee

The extracted $g$-factors are shown in Table I along with the 
measured values \cite{ga91} and results of a model calculation \cite{pi93}.
\begin{table}
\label{table:c60-mcd}
\caption{MCD response of the lowest four $^1$T$_{1u}$ states in C$_{60}$.  The
experimental data is from Ref. \cite{ga91}.  Our calculations are given
in the columns labelled TDDFT.  The
effective orbital $g$-factor is defined in Eq. (\ref{g}).
}
  \begin{tabular}{|cc|cc|ccc|}
  \colrule
 \multicolumn{2}{|c|}{Energy (eV)}      
&   \multicolumn{2}{|c|}{${\cal B}_n/{\cal D}_n$ } & 
\multicolumn{3}{|c|}{$g$}  \\
 Exp.  &  TDDFT   &   Exp. & TDDFT &  Exp. & TDDFT & Ref. \cite{pi93}  \\
\colrule
 3.8  & 3.5 &  100 &   64 & $-0.3\pm0.05$  &  $-0.97$ & $-1.0$   \\
 4.9  & 4.3 & -700 & -146 & $-0.55\pm0.15$ &  $-0.58$ & $-0.75$  \\
 6.0  & 5.3 &      &   66 &                &  $-0.20$ & $+0.12$  \\
      & 5.9 &      & -120 &                &  $+0.35$ &          \\
\colrule
\end{tabular}
\end{table}
As with the \calB$_n$ values, we see agreement on sign for the two
measured transitions.  However, only the upper 
transition has a magnitude consistent with experiment.  

\section{Concluding remarks}

We have shown that from a computational point of view, the real-time
method is a practical approach to calculate the MCD response in 
TDDFT.  In particular, the entire response in the energy region
of valence-electron excitations is obtained from a single calculation.
This allows one to use the sum rules, at least as a theoretical
tool, to understand the limitations with respect to the omission
of core electons from the dynamics.  It would be exceedingly 
challenging to 
ensure that the sum rules Eqs. (\ref{S0}) and
(\ref{S2}) are obeyed in 
formalisms that require the explicit construction of the excited state
spectrum.  

The violation of the sum rule Eq. (\ref{S2}) in the valence particle space
raises an issue that needs to be addressed in future work.  In Ref. 
\cite{ya98}, it was found that the violation of the dipole response in
TDDFT is largely justified.  The dynamic contribution of the core
electrons shifts oscillator strength down into the spectral region
of valence electrons, and this accounts physically for the increase
of the sum rule, calculated only with valence electrons employing the
nonlocal pseudopotential in the 
space of valence electron excitations.  Whether there is a related
mechanism to the decrease in the $I_2$ sum rule remains to 
be seen.  Also, the pseudopotential should in principle be corrected
for the gauge field associated with the magnetism, but that was not
done here.  It should be mentioned that these questions will also 
arise on calculations using the Projected Augmented Wave (PAW) 
method \cite{bl94}, since this also makes the Kohn-Sham operator nonlocal. 

It was also a surprise to us to find that the MCD response may
have an abnormal sign.  This goes against the picture of an 
electron being excited to a higher band of orbitals and there undergoing
circular motion in the sense given by the external magnetic field.
It might be that strong screening destroys the simple connection
to the expected classical oscillation picture.  This raises another
question for future work, to investigate in a general way the effects
of screening on the MCD.

Finally, we have not discussed here the sensitivity to specific
density functionals.  Although not reported, we have also carried out
the C$_{60}$ calculations with the LB94 functional \cite{le94}.  This
gave very similar results except for Rydberg transitions, which are
considerably shifted in energy, depending on the functional.  Since
the observables in MCD depend on currents, it might also be interesting
to investigate  the generalized TDDFT including current-current
interactions.

\section*{Acknowledgment}
This work was supported by the
National Science Foundation under Grant PHY-0835543 and by the 
DOE grant under grant DE-FG02-00ER41132.  Computations for $C_{60}$ were
carried out at the T2K supercomputer, University of Tsukuba,
and at the Supercomputer of Institute of Solid State Physics, University
of Tokyo.

\section*{Appendix}

In this Appendix we apply the real-time theory to a simple Hamiltonian,
a spinless electron in an anisotropic harmonic oscillator potential.
The model is completely solvable making it useful in checking the
coding and formulas for the TDDFT in a magnetic field.  

The Hamiltonian $H_0$ in Eq. (1) is taken as
\be
H_0 = {p^2\over 2 m} + {1\over 2} \sum_\alpha^3 m \omega_\alpha^2
r_\alpha^2
\ee
We label the eigenstates of $H_0$ by the number of excitation quanta
along each coordinate axis,  $|n_x n_y n_z\rangle$, and we set $m=e=\hbar=1$
in the equations below.  We take the oscillator frequencies $\omega_\alpha$
to be nondegenerate, so the MCD response will only have ${\cal B}$-type
contributions.  We first need the eigenstates in the presence of the 
magnetic field, expanded to first order in the field strength.  Taking the
magnetic field in the $z$-direction, the relevant perturbed orbitals are
\be
|000,B_z\rangle = |000\rangle -i s_0 |110\rangle
\ee
$$
|100,B_z\rangle = |100\rangle -i s_1 |010\rangle
$$
$$
|010,B_z\rangle = |010\rangle -i s_1 |100\rangle
$$
where
\be
s_{0,1} = {\mu_B B_z\over 2 }{\omega_y\mp \omega_x\over
(\omega_y\pm\omega_x)(\omega_x\omega_y)^{1/2}}.
\ee
The perturbed energies of the orbitals are not needed because that
perturbation is second order in $B_z$.

To get the real-time response $S^{(z)}_{xy}$, we multiply the ground state
wave function by the field $e^{ik \mu_y}$ and expand over the eigenstates,
to first order in $k$. The required matrix elements of the dipole operator between ground 
and excited states are
\be
\langle 100,B_z | \mu_x | 000, B_z\rangle=(2 \omega_x)^{-1/2}
\ee
$$ 
\langle 010,B_z | \mu_y | 000, B_z\rangle=(2 \omega_y)^{-1/2}
$$
$$
\langle 010,B_z | \mu_x | 000, B_z\rangle=-i\mu_B B_z (2
\omega_y)^{1/2}/(\omega_x^2-\omega_y^2)
$$
$$
\langle 100,B_z | \mu_y | 000, B_z\rangle=-i\mu_B B_z(2 \omega_x)^{1/2}/(\omega_x^2-\omega_y^2)
$$
The initial perturbed wave function is
\be
\label{initial}
|\Psi_{ky}(t=0)\rangle= | 000, B_z\rangle+i{k\over (2\omega_y)^{1/2}}| 010, B_z\rangle
+ \mu_B B_z k{(2 \omega_x)^{1/2}\over(\omega_x^2-\omega_y^2)} |100, B_z\rangle.
\ee
The time dependence is put in by multipling the excited states by
$e^{-i\omega_\alpha t}$.  The expectation value of $\mu_x$ may then be evaluated
as a function of time.  The result after some simplification is
\be
\label{rt-ho}
S^{(z)}_{xy} = {2 k \mu_B B_z \over \omega_x^2-\omega_y^2} (\cos \omega_x t
-\cos \omega_y t).
\ee
  The short-time response given by Eq. (27) may be verified by making a power
series expansion of the cosine functions in Eq. (\ref{rt-ho}).  Finally, the
evaluation of $R_{\rm MCD}$ by Eq. (20) may be verified by carrying out the
cosine Fourier transform, $\frac{2}{\pi}\int_0^\infty d t \cos \omega t\, \cos
\omega_0 t = \delta(\omega-\omega_0)$.  
Putting
in all three magnetic moment directions, the result is
\be
R_{\rm MCD} = -{1\over 3}{1\over \mu_B B} \sum_{\beta\ne\alpha} \delta(E-\omega_\alpha)
{2\over \omega_\alpha^2-\omega_\beta^2}
\ee
It may be easily verified that $R_{\rm MCD}$ satisfies the two sum rules,
Eq. (25) and (26).

\end{document}